\documentclass[11pt,a4paper]{article}

\usepackage{jheppub}
\usepackage{enumitem}   
\usepackage[english]{babel}
\usepackage{url}
\usepackage{mathtools}
\usepackage{bbold}
\usepackage{slashed}
\usepackage{multirow}
\usepackage{slashed}
\usepackage{multirow}
\usepackage[usenames,dvipsnames,svgnames]{xcolor}
\usepackage{appendix}

\newcommand{\be}{\begin{equation}}
\newcommand{\ee}{\end{equation}}

\def\be{\begin{equation}}
\def\ee{\end{equation}}

\begin{document}



\title{Axionic domain walls at Pulsar Timing Arrays: QCD bias and particle friction}

\author[a]{Simone Blasi,}
\author[a]{Alberto Mariotti,}
\author[a]{A\"aron Rase,}
\author[b,c]{Alexander Sevrin}

\affiliation[a]{Theoretische Natuurkunde and IIHE/ELEM, Vrije Universiteit Brussel, \& The  International Solvay Institutes, Pleinlaan 2, B-1050 Brussels, Belgium}
\affiliation[b]{Theoretische Natuurkunde, Vrije Universiteit Brussel, \& The  International Solvay Institutes, Pleinlaan 2, B-1050 Brussels, Belgium}
\affiliation[c]{Universiteit Antwerpen, Prinsstraat 13, 2000 Antwerpen, Belgium}

\emailAdd{simone.blasi@vub.be}
\emailAdd{alberto.mariotti@vub.be}
\emailAdd{aaron.rase@vub.be}
\emailAdd{alexandre.sevrin@vub.be}



\abstract{The recent results from the Pulsar Timing Array (PTA) collaborations show the first evidence for the detection of a stochastic background of gravitational waves at the nHz frequencies. This discovery has profound implications for the physics of both the late and the early Universe. In fact, together with the possible interpretation in terms of super massive black hole binaries, many sources in the early Universe can provide viable explanations as well. In this paper, we study the gravitational wave background sourced by a network of axion--like--particle (ALP) domain walls at temperatures around the QCD crossover, where the QCD--induced potential provides the necessary bias to annihilate the network. Remarkably, this implies a peak amplitude at frequencies around the sensitivity range of PTAs. We extend previous analysis by taking into account the unavoidable friction on the network stemming from the topological coupling of the ALP to QCD in terms of gluon and pion reflection off the domain walls at high and low temperatures, respectively. We identify the regions of parameter space where the network annihilates in the scaling regime ensuring compatibility with the PTA results, as well as those where friction can be important and a more detailed study around the QCD crossover is required.}


\maketitle

\section{Introduction}
The PTA consortium has very recently reported the first positive evidence of a stochastic gravitational wave background (SGWB)
around the nHz frequency
\cite{NANOGrav:2023hvm,NANOGrav:2023hde,NANOGrav:2023gor,Antoniadis:2023lym,Antoniadis:2023ott,Smarra:2023ljf,Zic:2023gta,Reardon:2023gzh,Xu:2023wog} (see \cite{Perera:2019sca,Antoniadis:2022pcn,NANOGrav:2020bcs,Chen:2021rqp,Goncharov:2021oub} for previous PTA results).
One possibility is that
this signal is of astrophysical origin and that it consists of a superposition coming from super massive black hole binaries\,\cite{Burke-Spolaor:2018bvk}.
However, it is interesting to explore the possibility that the background observed at PTAs is actually of cosmological origin.
In this case, this discovery would provide unique new information about the early history of our Universe.
Several works have recently appeared, with the aim of interpreting this new SGWB signal at PTAs\,\cite{Han:2023olf,Guo:2023hyp,Megias:2023kiy,Fujikura:2023lkn,Yang:2023aak,Kitajima:2023cek,Bai:2023cqj,Zu:2023olm,Vagnozzi:2023lwo,Lambiase:2023pxd,Ellis:2023dgf,Li:2023yaj,Franciolini:2023wjm,Shen:2023pan,Ellis:2023tsl,Franciolini:2023pbf,Wang:2023len,Ghoshal:2023fhh,King:2023cgv},
for previous studies see e.g.\,\cite{Blasi:2020mfx,Ellis:2020ena,Buchmuller:2020lbh,Bian:2020urb,Blanco-Pillado:2021ygr,Sakharov:2021dim,Bian:2022tju,Chen:2022azo,Wang:2022rjz,Ferreira:2022zzo,Ramberg:2022irf,Bringmann:2023opz,Madge:2023cak}.
The NANOGrav collaboration has already considered several new physics interpretations of their 
dataset as well\,\cite{NANOGrav:2023hvm}.

In this paper, we focus on the scenario where the detected SGWB is sourced by domain walls (DWs), two--dimensional topological defects that can arise during a phase transition involving the spontaneous breakdown of a discrete symmetry\,\cite{Kibble:1976sj,Zeldovich:1974uw}. 
DW networks are predicted in many scenarios of
physics Beyond the Standard Model (BSM) and their
signatures haven been recently investigated in 
several works \cite{Higaki:2016jjh,Gelmini:2021yzu,Gelmini:2020bqg,Gelmini:2022nim,Craig:2020bnv,ZambujalFerreira:2021cte,
Babichev:2021uvl,Barman:2022yos,Wu:2022tpe,Borah:2022wdy,Wu:2022stu,Ferreira:2022zzo,Fornal:2022qim,Pujolas:2022qvs,Gelmini:2023ngs,Fornal:2023hri,Bosch:2023spa,Blasi:2023rqi,
Bian:2020urb,Wang:2022rjz,Sakharov:2021dim,Blasi:2022woz,Blasi:2022ayo,Madge:2023cak}
(see \cite{Vilenkin:2000jqa,Saikawa:2017hiv} for reviews on domain walls).

A DW network in the early Universe can be a powerful source of gravitational waves, as shown quantitatively in numerical simulations of the corresponding field theory\,\cite{Hiramatsu:2010yz,Hiramatsu:2012sc,Hiramatsu:2013qaa,Saikawa:2017hiv}.
The NANOGrav collaboration has in fact interpreted this signal as coming from a DW network in the scaling regime\,\cite{NANOGrav:2023hvm}, identifying the ranges of temperature and energy density of the network compatible with the data. We will employ their results in our analysis.

A scenario where DWs arise naturally is in models of axion--like--particles (ALPs). ALPs are generalizations of the Peccei Quinn QCD axion solving the strong CP problem \cite{Peccei:1977hh,Peccei:1977ur,Weinberg:1977ma,Wilczek:1977pj}, and they are common in BSM physics, see e.g. \cite{Svrcek:2006yi,Arvanitaki:2009fg,
Holdom:1982ex,Holdom:1985vx,Flynn:1987rs,Rubakov:1997vp,Choi:1998ep,Berezhiani:2000gh,Hook:2014cda,Fukuda:2015ana,Dimopoulos:2016lvn,Agrawal:2017ksf,Gaillard:2018xgk,
Preskill:1982cy,Abbott:1982af,Dine:1981rt,Dine:1982ah,Arias:2012az}.

DWs arise naturally in ALP models where the discrete subgroup of the $U(1)$ which is preserved by the anomaly undergoes spontaneous symmetry breaking. These DWs can be topologically stable and eventually dominate the energy density of the Universe \cite{Zeldovich:1974uw,Sikivie:1982qv}, in conflict with standard cosmology.
This problem can be avoided by introducing a (small) explicit breaking of the discrete symmetry (the so--called \emph{bias}), leading to the annihilation of the network \cite{Sikivie:1982qv,Gelmini:1988sf}.
The size of the bias is generally a new independent parameter, which determines the phenomenology of the network and in particular the emitted SGWB.

As we shall discuss, a natural scenario to motivate the signal from ALP DWs at the PTA frequencies is the one where the ALP couples to QCD, as already suggested e.g. in \cite{Preskill:1991kd,Hiramatsu:2012sc,Ferreira:2022zzo,Blasi:2022ayo,Madge:2023cak}, so that the QCD--induced potential itself acts as a bias.
In this case, however, 
friction from the QCD sector in the thermal plasma is inevitable. 
Determining the impact of this friction force for the ALP domain wall interpretation of the PTA data is the main goal of our study (see\,\cite{NAKAYAMA2017500,Huang:1985tt,Abel:1995wk,Vilenkin:2000jqa,Blasi:2022ayo} for previous studies of friction effects on DW networks).
In fact, when friction dominates the network departs from scaling and the corresponding SGWB can significantly change with respect to the one observed in numerical simulations\,\cite{Hiramatsu:2010yz,Hiramatsu:2012sc,Hiramatsu:2013qaa,Saikawa:2017hiv} where plasma effects are not included.

Our main result is that friction acting on the DWs from the QCD sector can in general be relevant for a significant part of the model parameter space capable of reproducing the NANOGrav signal, 
and that a departure from the scaling regime (on which the PTA interpretation is based) is possible, even though a more detailed analysis around the QCD crossover is required to completely settle the issue. We also find parameter space compatible with the SGWB at PTAs where friction can be safely neglected. 

Note that our study focuses on a minimal realization of 
ALP DWs with QCD-induced bias, where the heavy ALP cannot be the QCD axion. Nevertheless our analysis on friction effects can apply equally to non-minimal scenarios where the heavy axion is actually
the one responsible to solve the strong CP problem, see e.g.
\cite{Higaki:2016jjh,Long:2018nsl}.

Additionally, while the study presented in this paper focuses on the unavoidable coupling with QCD, 
friction could also impact the DW network if other ALP-SM (model--dependent) couplings are present.
For instance, in Ref.\,\cite{Blasi:2022ayo} it was shown that if the ALP couples to muons\footnote{The other SM fermions are either too weakly coupled (the light ones), or generally too heavy to be abundant in the plasma at the QCD crossover.} the resulting friction can actually play a role at PTA frequencies provided that this interaction has the right strength.

The structure of the paper is as follows.
In the next section  we review the basics of ALP domain walls and SGWBs.
Then, in Section \ref{QCDbias} we show that ALP DWs whose bias is induced by QCD are a natural candidate to explain the PTA signal, by comparison with the model--independent analysis in\,\cite{NANOGrav:2023hvm}.
In Section \ref{frictionQCD} we study the impact of friction from the QCD sector, analyzing the contribution from gluons and pions at high and low temperatures.
In section \ref{sec:results} we present our results showing the relevance of friction for the ALP parameter space.

\section{ALP domain walls}
\label{DWatPTA}
We consider domain walls that generically arise in ALP models. 
We introduce our notation and setup. We start off by considering a dark QCD non abelian group $SU(N)$ and a $U(1)$ Peccei--Quinn symmetry anomalous under $SU(N)$.
The corresponding pseudo--Nambu--Goldstone boson, $a$, plays the role of the ALP in our analysis. The Lagrangian for the ALP contains the following term
\begin{equation}
    \mathcal{L}_a \supset \frac{\alpha_d}{4\pi} \frac{N_d}{v} a \,G_d \widetilde G_d,
\end{equation}
where $\alpha_d$ is the dark gauge coupling constant, $v$ is the $U(1)$ breaking vev, and $G_d$ is the dark gauge boson field strength, which is contracted with its dual $\tilde G_d$. In terms of the quantities above it is useful to define the ALP decay constant $f_a$ as
\begin{equation}
    f_a = \frac{v}{N_{\rm DW}}, \quad N_{\rm DW} = 2 N_d.
\end{equation}
In terms of this quantity the Lagrangian is defined as usual as 
\begin{equation}
\mathcal{L}_a \supset \frac{\alpha_d}{8\pi} \frac{a}{f_a} G_d \tilde G_d.
\end{equation}
For $N_d = 1/2$ the domain wall number is one and the vacuum manifold for the ALP potential induced by the dark gauge theory is trivial, namely it contains only one minimum as $a=0$ and $a=2\pi f_a = 2 \pi v$ are to be identified. For $N_{\rm DW} > 1$ however the vacuum consists of disconnected points, and stable domain wall solutions exist interpolating between neighboring minima.

The most important features of the ALP potential are captured by the following structure which encodes the discrete $Z_{N_{\rm DW}}$ of the ALP Lagrangian with respect to the dark sector:
\begin{equation}
V_d(a) = m_a^2 f_a^2 \left(1 - \,{\rm cos}\left( \frac{a}{f_a}\right) \right),
\end{equation}
where $m_a$ is the ALP mass.
The ALP potential is defined in the range $a \in [0, 2 \pi v) = [0, 2 \pi N_{\rm DW} f_a)$ and it then supports $N_{\rm DW} -1$ degenerate and inequivalent minima. The simple cosine potential allows to obtain analytical domain wall solutions,
\begin{equation}
\label{eq:DWprofile}
    a(z) = \left[ 2 \pi k + 4 {\rm arctan}(e^{m_a z}) \right] f_a, \quad k=0,1, \dots N_{\rm DW}-1,
\end{equation}
with energy per unit surface (tension) given by
\begin{equation}
    \sigma_{\rm DW} = 8 m_a f_a^2.
\end{equation}

In general, the ALP potential could differ significantly from the cosine shape if additional (pseudo) Nambu--Goldstone bosons exist below the dark QCD confinement scale, as it is the case for the QCD axion, see e.g.\,\cite{Villadoro}.
In any case, the profiles above can still capture most of the relevant physics of the ALP domain wall solution\,\footnote{Note however that while the cosine potential predicts vanishing ALP self--reflection off the ALP domain wall, a QCD--like potential was shown to not maintain this property\,\,\cite{Blasi:2022ayo}.}.

\paragraph{Cosmology evolution of DW and SGWB spectrum}
We assume that the PQ breaking scale $v$ is smaller than the reheating temperature
and we focus on the case in which the domain wall number is larger than one, so that the DW network is stable \cite{Vilenkin:1982ks,Sikivie:1982qv}.\footnote{For $N_{\text{DW}}$=1 the network is unstable and decays soon after formation\,\cite{Vilenkin:1982ks,Barr:1986hs,Shellard:1986in,BARR1987591,Chang:1998tb}.}
At the scale of the PQ phase transition global strings associated to $U(1)$ are formed according to the Kibble mechanism \cite{Kibble:1976sj}, see\,\cite{Gorghetto:2018myk,Hindmarsh:2019csc,Gorghetto:2020qws,Gorghetto:2022ikz} for recent work.
Subsequently, when the ALP potential becomes cosmologically relevant, that is when $H(T_{\text{f}}) \sim m_a$, domain walls can be considered formed. 
The precise relation to determine $T_{\text{f}}$ should take into account the temperature dependence of the axion potential generated by the dark-QCD sector.
For an ALP decay constant below the Planck scale, one can show that $T_{\text{f}} \gtrsim \sqrt{m_a f_a}$, so we consider $\sqrt{m_a f_a}$  as an estimate of $T_{\text{f}}$ that will anyway play no important role in our study.

Soon after DW formation, the energy density of the resulting string--wall hybrid network is soon dominated by the walls\,\cite{Vilenkin:2000jqa,Hiramatsu:2012sc}.
The DWs reach then the so-called scaling regime where the energy density of the DW network redshifts as $\rho_{\text{DW}}\sim \sigma H$, corresponding to $\mathcal{O}(1)$ DWs per Hubble patch and mildly relativistic average velocity\,\cite{Ryden:1989vj,Hindmarsh:1996xv,Garagounis:2002kt,Oliveira:2004he,Avelino:2005pe,Leite:2011sc}.
 
A scaling network of DWs will eventually dominate the energy density of the Universe, in contrast to standard cosmology \cite{Zeldovich:1974uw,Sikivie:1982qv,Vilenkin:1981zs}.
This occurs when
$\rho_{\text{DW}} \sim 3 H^2 M_{\text{Pl}}^2$, or in terms of temperature
\begin{equation}
\label{Tdom}
    T_{\text{dom}} \simeq 14 ~\text{MeV} \left(\frac{\sigma_{\rm DW}^{1/3}}{100~ \text{TeV}} \right)^{3/2} \left(  \frac{g_*}{10} \right)^{-1/4},
\end{equation}
where we have assumed radiation domination, and we have used $\rho_{\text{DW}}= 2\sigma H \mathcal{A}$ 
with $\mathcal{A}=0.8$ from numerical simulations\,\cite{Hiramatsu:2010yz,Hiramatsu:2012sc,Hiramatsu:2013qaa,Saikawa:2017hiv}.

In order to collapse the DW network before domination, one may add a bias potential $\Delta V$ that breaks explicitly the $Z_{N_{\rm DW}}$ symmetry. We shall then define $T_{\ast}$ as the the annihilation temperature, where $T_{\ast}> T_{\text{dom}}$ for consistency. 

At the time of annihilation, the DW energy density normalized to radiation is given by
\begin{equation}
\label{eq:alphastar}
    \alpha_{\ast} = \frac{\rho_{\text{DW}}}{3 H^2 M_{Pl}^2}  \simeq 0.02
    \left(\frac{\sigma_{\rm DW}^{1/3}}{100~ \text{TeV}} \right)^{3}
    \left(\frac{T_{*}}{100~\text{MeV}} \right)^{-2}
    \left(  \frac{g_*}{10} \right)^{-1/2}.
\end{equation}
This definition of $\alpha_{\ast}$ is inspired by analogous studies of first order phase transitions, see e.g.\,\cite{Espinosa:2010hh}, and as we shall see 
it is directly related to the strength of the GW emission.

The annihilation temperature may be estimated by balancing the curvature pressure with the energy difference induced by the bias, namely $\Delta V \sim \sigma_{\rm DW}/R \sim \rho_{\text{DW}}$, where $R$ is the correlation length of the network.
One finds
\begin{equation}
\label{Tannsimple}
T_{*} \simeq 270 ~\text{MeV} 
\left(\frac{\sigma_{\rm DW}^{1/3}}{100~ \text{TeV}} \right)^{-3/2}
\left(\frac{\Delta V^{1/4}}{100~\text{MeV}} \right)^{2}
\left(  \frac{g_*}{10} \right)^{-1/4},
\end{equation}
where we used the condition $\Delta V= \mathcal{C}_{\text{ann}}~\rho_{\text{DW}}$
with $\mathcal{C}_{\text{ann}} \simeq 2$ from numerical simulations
\cite{
Hiramatsu:2010yz,Hiramatsu:2012sc,Hiramatsu:2013qaa,Saikawa:2017hiv}.
The bias is in principle a free parameter that should be added to the model, and the phenomenology of the DW network can change drastically depending on its size, 
In general, one can expect this to come from quantum gravity effects that make the starting $U(1)$ global symmetry only approximate \cite{Barr:1992qq,Kamionkowski:1992mf,Holman:1992us,Berezhiani:1992pq,Ghigna:1992iv,Senjanovic:1993uz,Dobrescu:1996jp,Banks:2010zn}.
On the other hand, the size of the bias can be predicted if it is dynamically generated. In the next section we will explore the possibility that such bias is in fact generated by QCD.

The DW network in the scaling regime has been proven by numerical simulations \cite{Hiramatsu:2012sc,Hiramatsu:2013qaa,Saikawa:2017hiv} to generate a large SGWB $\Omega_{\text{gw}}(f)$ with broken power law in frequency.
The signal is dominated by the last moment of emission, so it depends explicitly on $T_{\ast}$.
The signal redshifted today has the form
\begin{equation}
\Omega_\text{gw}(T_{\text{ann}},f) = \Omega_{\text{peak}}\times
\begin{cases}
      \left(\frac{f}{f_\text{peak}}\right)^{3} & \text{if $f \leq f_\text{peak}$}\\
      \left(\frac{f}{f_\text{peak}}\right)^{-1} & \text{if $f>f_\text{peak}$}
    \end{cases}
\end{equation}
with
\begin{eqnarray}
\label{GWfromDW}
    \Omega_{\text{peak}} &
\simeq &
 1.64\times 10^{-6} \left( \frac{\Tilde{\epsilon}_\text{gw}}{0.7} \right)
 \left(\frac{\mathcal{A}}{0.8}
 \right)^{2}
 \left(\frac{g_*(T)}{10}\right)\left(\frac{g_{*s}(T)}{10}\right)^{-4/3}\left(\frac{T_\text{dom}}{T_{\text{ann}}}\right)^4 \\
f_{\text{peak}} & \simeq &
1.15\times 10^{-9} ~ \text{Hz}\left(\frac{g_*(T)}{10}\right)^{1/2}\left(\frac{g_{*s}(T)}{10}\right)^{-1/3}\left(\frac{T_{\text{ann}}}{10\ \text{MeV}}\right),
\end{eqnarray}
where we have normalized the numerical coefficient to the values obtained in numerical simulations \cite{Hiramatsu:2010yz,Hiramatsu:2012sc,Hiramatsu:2013qaa,Saikawa:2017hiv}.
The SGWB spectrum of DW is determined by two parameters, the tension (see Eq.\,\eqref{Tdom}) and $T_{\ast}$ \footnote{There is also a mild dependence on $N_{\text{DW}}$ which is encoded in O(1) modifications of the numerical coefficient $\mathcal{A}$ \cite{Kawasaki:2014sqa}.}.
Eq.\,\eqref{GWfromDW} shows that the later the DW network annihilates, the larger the GW 
signal is.
As we can see, the best $T_{\ast}$ for PTA frequencies is in the ballpark of the QCD scale.

\section{The QCD potential as the natural bias for DWs at PTAs}
\label{QCDbias}

Let us now consider the effect of the QCD--induced potential on the ALP model illustrated above. 
This comes from the anomalous coupling between the ALP and the gluons,
\begin{equation}
\label{Glugluinte}
    \mathcal{L}_a \supset \frac{\alpha_s}{4 \pi} \frac{N_c}{v} G \tilde G,
\end{equation}
where $N_c$ is the color anomaly from fermions charged under QCD. In general, $N_c$ and $N_d$ are two independent numbers. Whenever these numbers are not coprime, the degeneracy in the vacuum manifold is lifted and domain walls become metastable.

The contribution to the ALP potential from QCD at low energy can be captured within to chiral perturbation theory, see e.g. \cite{Villadoro,DiLuzio:2020wdo}. One finds the following potential for the ALP--pion system:
\be
\label{villagold}
V(a, \pi_0) = - \frac{f_\pi^2 m_\pi^2 }{m_u + m_d}
\left[ m_u \,{\rm cos}\left( \frac{a}{2 f_a^\prime} - \frac{\pi_0}{f_\pi}\right) + m_d \,{\rm cos}\left( \frac{a}{2 f_a^\prime} + \frac{\pi_0}{f_\pi}\right)\right].
\ee
and 
\begin{equation}
    f_a^\prime \equiv \frac{N_d}{N_c}f_a.
\end{equation}
Notice that since $f_a^\prime \neq f_a$, the periodicity of the QCD potential is generally misaligned with respect to the one of the dark--QCD potential.
The interactions in \eqref{villagold} follow from an ALP--dependent rephasing of the light up and down quarks that removes the ALP from the topological term, $q \rightarrow q \,{\rm exp}(i \gamma_5 \frac{a}{2f_a} Q_a)$ and $Q_a$ proportional to the identity ($\rm{Tr}\,Q_a = 1$).

Assuming that the QCD contribution is very small compared to the dark QCD one ($m_{\pi} f_{\pi} \ll m_a f_a$),
the size of the bias is generically given by
$|\Delta V_k| \sim m_\pi^2 f_\pi^2$.
However, when the two sectors are almost aligned, for instance when $N_c/N_d = 1 + \epsilon$ with $\epsilon \ll 1$, the bias can become parameterically small:
\begin{equation}
\label{eq:bias}
    |\Delta V_k| \sim \epsilon^2 m_\pi^2 f_\pi^2
\end{equation}
and the life time of the network may be parameterically enhanced. In our analysis we will keep $\epsilon$ as a free parameter, keeping in mind that a scenario with $\epsilon \ll 1$ requires somewhat large or fine--tuned values of $N_c$ and $N_d$.

\medskip

Let us now turn to discuss how temperature corrections modify the size of the QCD--induced ALP potential. At very high temperatures above QCD confinement, 
the ALP potential is expected to behave as
\cite{Gross:1980br,Borsanyi:2016ksw,Villadoro,DiLuzio:2020wdo,OHare:2021zrq}
\begin{equation}
\label{eq:highTpot}
    V(a; T) =  \chi(T) \left[ 1-  {\rm cos}\left(\frac{a}{f_a^\prime}\right)\right] = \chi_0 \left(\frac{T}{150 \, {\rm MeV}}\right)^{-n} \left[ 1-  {\rm cos}\left(\frac{a}{f_a^\prime}\right)\right],
\end{equation}
with $n \simeq 7$ and $\chi_0^{1/4} \simeq 75.6$ MeV, even though some uncertainty on these parameter still remains (see e.g.\,\cite{OHare:2021zrq} and references therein).
Similarly to the low--temperature case, some approximate alignment between QCD and the dark QCD can lead to a parametric suppression of the natural bias
$|\Delta V_k|(T) \sim \chi(T)$ to $|\Delta V_k|(T) \sim \epsilon^2 \chi(T)$.

\begin{figure}[t]
    \centering
    \includegraphics{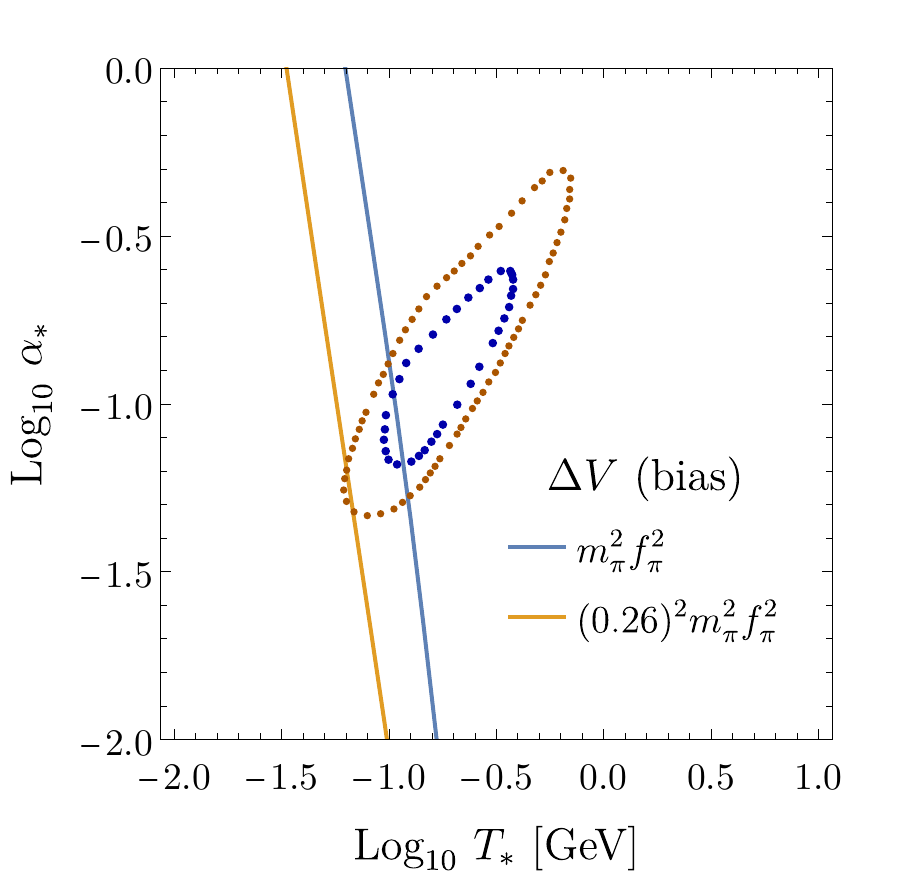}
    \caption{  
    One and two sigma contours for the DW interpretation of the signal as provided by the \cite{NANOGrav:2023hvm} collaboration (blue and yellow dots).
    The prediction of a DW network with QCD induced bias ($\Delta V \sim \epsilon^2 m_{\pi}^2 f_{\pi}^2$)
    are displayed as lines with varying DW tension.}
    \label{fig:biasvsfit}
\end{figure}

From this simple estimate we can already draw some conclusions in the light of the recent PTA results.
In Ref.\,\cite{NANOGrav:2023hvm}, the collaboration has performed a bayesian analysis on the NANOGrav data for the DW interpretation. 
The results were displayed in a two dimensional plane of $T_{*}$ vs $\alpha_{*}$ as 1 and 2 sigma contours, as shown in Figure \ref{fig:biasvsfit}, for the case of DWs as the only source contributing to the GW signal.

In order to compare the NANOGrav contours with the scenario we are discussing, we can first use equations \eqref{eq:alphastar} and \eqref{Tannsimple} to relate directly the fraction of energy density to the annihilation temperature, for a given bias potential,
\begin{equation}
    \alpha_{*} \simeq 0.15 \left(\frac{\Delta V^{1/4}}{100~\text{MeV}} \right)^{4}
    \left(\frac{T_{*}}{100~\text{MeV}} \right)^{-4}
     \left(  \frac{g_*}{10} \right)^{-1}.
     \end{equation}
Plugging in $\Delta V \sim \epsilon^2 m_{\pi}^2 f_{\pi}^2$ in the equation above we obtain a line in the $T_{*}$ vs $\alpha_{*}$ plane. Notice that each point on this line corresponds to a specific domain wall tension.

We display in Figure \ref{fig:biasvsfit}
the lines of two representative cases for $\epsilon$, showing how the QCD--induced bias can accommodate the NANOGrav data.
Notice that since $\epsilon \leq 1$ in this minimal realization, we cannot access the entire region favoured by the NANOGrav analysis in this model.

The results from the NANOGrav collaboration are obtained in a model--independent way assuming that the DW network is in the scaling regime at annihilation. In the following section, we shall investigate whether this assumption is compatible with the natural QCD bias.

\section{The impact of friction from QCD}
\label{frictionQCD}

In this section we study the friction acting on the domain wall as a consequence of the reflection of particles in the plasma.
In particular, we are interested in friction effects close to the annihilation temperature of the network in the range relevant for Fig.\,\ref{fig:biasvsfit}.
If friction dominates before annihilation, the DW network is not in scaling and the predictions for the GW spectrum can change significantly\,\cite{NAKAYAMA2017500,Blasi:2022ayo}, possibly jeopardizing the PTA interpretation.
The irreducible friction on ALP domain walls in scenarios which the QCD bias comes from gluons and in general from hadrons at low temperatures.

The effect of friction is usually parameterized by defining a friction length $\ell_f$ which feeds in the total damping scale of the network $\ell_d$ as 
\begin{equation}
\label{eq:damping}
    \frac{1}{\ell_d} = 3H + \frac{1}{\ell_f},
\end{equation}
where $H$ is the Hubble parameter and 
\begin{equation}
    \frac{1}{\ell_f} = \frac{\Delta P}{v_w \sigma_{\rm DW}}.
\end{equation}
Here $v_w$ is the average velocity of the network, and $\Delta P$ is the pressure on the wall from interactions with the plasma. The definition above takes into account that one generally expects $\Delta P \propto v_w$, at least for moderate velocities. 

The pressure can be computed from an integral over the particle thermal distribution involving the reflection coefficient\,\cite{Arnold:1993wc,Blasi:2022ayo},  $\mathcal{R}(p_z)$,
\begin{equation}
\label{eq:pressurea}
    \Delta P = \frac{2 g}{(2\pi)^2}
    \int_0^\infty \text{d}p_z p_z^2 \mathcal{R}(p_z)
    \frac{1}{\beta \gamma a}
    \left[ \,
    2 \beta \gamma p_z v - \text{log}\left(\frac{f(-v)}{f(v)}\right)\right]\bigg|_{E=\sqrt{p_z^2+m^2}},
\end{equation}
where $a=\pm1$ for FD or BE statistics respectively, $g$ counts the number of d.o.f,
and $f(v)$ is the thermal distribution in the wall rest frame
    \begin{equation}
    f(v) = \frac{g}{e^{\gamma(v) \beta(E+ p_z v)}\pm1}.
\end{equation}
The reflection coefficient $\mathcal{R}(p_z)$ can be computed by solving the quantum mechanical reflection
for a particle scattering off the ALP wall.
For temperatures such that $1/\ell_{f} \gtrsim 3H$ the DW network deviates from scaling and enters a friction dominated regime, where one expects suppressed GW signals\,\cite{NAKAYAMA2017500,Blasi:2022ayo}.

The temperatures of interest are again in the ballpark of the QCD crossover.  
We face the problem following 
a very simple strategy
\footnote{This should be seen as a first approximation to the problem of hadrons scattering off axionic DWs. A more detailed analysis, for instance along the lines used for thermal axion production in\,\cite{Notari:2022zxo,DEramo:2021lgb}, is left for future work.}
and we study two regimes at temperatures above and below the QCD scale:
\begin{itemize}
\item For $T\gtrsim2$ GeV we compute friction by considering the scattering of gluons off the ALP DW through the interaction \eqref{Glugluinte}. 
\item For $T \lesssim 60$ MeV we employ chiral perturbation theory and the main source of friction is induced by scattering of pions off the ALP DW.
\end{itemize}
Even if we cannot compute the friction in the intermediate temperature regime, we will be able to draw interesting conclusions concerning the DW dynamics around the QCD crossover.

\subsection{Friction from gluons}
At high temperatures, the contribution to $\Delta P$ comes from gluons reflecting off the ALP domain wall. We stress that this effect is unavoidable in the scenario in which QCD provides the bias collapsing the network.

In the simplified picture of friction as coming from one--to--one particle reflection off the ALP wall, we can neglect as a first approximation the gluon self interactions and work at the linear order in the field fluctuations. In this limit, one recovers independent abelian equations of motion for each gluon degree of freedom. 
Additionally, when the ALP wall comes from the simplest cosine potential, a reasonable approximation for the reflection coefficient can be obtained analytically\,\cite{Ganoulis:1986rd} (see also \,\cite{Huang:1985tt}).
The reflection probability for a negative helicity gluon is given by 
\begin{equation}
\label{eq:ganoulis}
    R^-(\rho) = \frac{1 + \,{\rm cos}(\pi \sqrt{1+\beta \rho})}{{\rm cos}(\pi \sqrt{1+\beta \rho}) + \,{\rm cosh}(4\pi \rho)},
\end{equation}
where $\rho = p_z/m_a$, with $p_z$ the gluon momentum in the direction orthogonal to the wall, and 
\begin{equation}
    \beta = \frac{4 N_c}{\pi N_d} \alpha_s.
\end{equation}
The reflection for positive helicity gives a quantitatively very similar result as in \eqref{eq:ganoulis} in the case of interest. At momenta much below the inverse width of the domain wall $\sim m_a$ (which also sets the height of the potential/well seen by the gluon depending on the helicity) particles have a finite probability of being reflected $\propto \beta^2$.

The pressure induced by this type of interaction may be computed according to \eqref{eq:pressurea}
and grows with the temperature as
\begin{equation}
\label{eq:gluonlargema}
    \Delta P = v_w \cdot g \frac{3}{32 \pi^2} \beta^2 T^4,
    \quad m_a \gg T,
\end{equation}
where $g=2 \cdot 8$ for gluons. For temperatures $T \gtrsim m_a$ an even larger fraction of particles is simply transmitted and the $\propto T^4$ behavior is tamed to a much slower increase $\propto T$,
\begin{equation}
\label{eq:gluonhighT}
    \Delta P = 2 \cdot 10^{-5} v_w \cdot g \,\beta^2 m_a^3 \, T,
    \quad T \gg m_a.
\end{equation}
Our calculation of the friction at low temperatures is of course limited by QCD becoming non--perturbative. In our analysis we will push this description down to $T \gtrsim 2\, \rm GeV$, keeping in mind that corrections should be expected at the low end of this region.

Notice also that there is a source of model dependence given by the ratio of the QCD and dark QCD anomaly, $N_c/N_d$. Clearly, when $N_c \rightarrow 0$ the ALP is decoupled from QCD and indeed \eqref{eq:gluonlargema} and \eqref{eq:gluonhighT} yield a vanishing contribution. In the following we shall take $N_c/N_d =\mathcal{O}(1)$, keeping in mind that decoupling the ALP from QCD is anyway not compatible with the generation of the required bias term.

Let us also notice that the pressure from gluon reflection is generally much bigger than the high--temperature bias induced by the QCD instantons in \eqref{eq:highTpot}.
This crucially implies that the network can consistently reach a friction--dominated regime well before annihilation begins. 

\subsection{Friction from pions}
\label{sec:frictionfrompions}
In order to evaluate the friction from pions we refer to chiral perturbation theory and consider the potential
in \eqref{villagold}.
Let us first stress that in the scenario of interest where the QCD contribution is not aligned with the potential induced by the dark QCD, the pion mass will change in the different vacua. This simply signals that the degeneracy has been indeed removed. Taking this into account, pressure from the pions  is expected to be of the same order of the potential bias.

However, this pressure is not what we are interested in, as this would only determine the terminal velocity during the domain wall network collapse. Instead, the question we wish to address is whether the ALP interaction with pions could in principle turn a scaling network (where the bias is by definition irrelevant) into a friction--dominated evolution. 

To determine this, we shall evaluate the pion pressure in a system in which the QCD potential and the dark--QCD potential are aligned. The pressure comes then from pion reflection off the domain wall, with the pion mass being the same on both sides. This has to be interpreted as a lower bound on the pion pressure.

We then set $f_a = f_a^\prime$ only for this specific calculation, and we additionally assume a large hierarchy between the ALP and the pion mass, $m_a \gg m_\pi$.
This allows us to neglect the backreaction of the pion on the ALP domain wall solution following its own dark--QCD potential. We then set the ALP to its $z$--dependent background given in \eqref{eq:DWprofile}, and study the motion of the $\pi_0$ around it.

First of all, we have to take into account that the ALP background induces a z-dependent background on the $\pi_0$ as well, given as the solution of
\be
-\pi_b^{\prime \prime}(z) + \frac{\partial V(a(z),\pi_0)}{\partial \pi_0}\big|_{\pi_0 = \pi_b(z)} = 0.
\ee
The structure of the vacua implies that the profile for $\pi_b$ is such that $\pi_b(-\infty) = 0$
(where $a(-\infty) = 0$), and $\pi_b(+\infty)/f_\pi = \pi$ (where $a(-\infty)/2f_a = \pi$).

In the assumed hierarchy $m_a \gg m_\pi$, the ALP background varies on scales much shorter than the inverse pion mass. Thus, the equation above can be split on the two sides of the ALP wall as
\be
\label{eq:pibminus}
-\frac{\pi_b^{\prime \prime}(z)}{f_\pi} + m_\pi^2 \,{\rm sin}\left( \frac{\pi_b(z)}{f_\pi} \right) = 0, \quad z < 0,
\ee
and 
\be
\label{eq:pibplus}
-\frac{\pi_b^{\prime \prime}(z)}{f_\pi} - m_\pi^2 \,{\rm sin}\left( \frac{\pi_b(z)}{f_\pi} \right) = 0, \quad z > 0,
\ee
where we have used the asymptotics for $a(z)$.
Once the solution for $\pi_b(z^-)$ is obtained for $z<0$, the solution for $z>0$ is given by
\be
\pi_b(z^+) = \pi - \pi_b(-z^+). 
\ee
As we can see, the equations of motion for $\pi_b$ are similar to the ones for the sine--Gordon model, where the potential would instead be $\propto {\rm cos}(\phi)$ for both positive and negative $z$. Our case is however distinct to the sine--Gordon as both the equation of motion and the boundary conditions for $\pi_b$ are different. Contrary to the sine--Gordon model where particle excitations are exactly (self) reflectionless, we then generically expect a non--zero reflection coefficient\,\footnote{Notice that in the ALP decoupling limit, $f_a^\prime \rightarrow \infty$ with $f_a$ fixed, the influence of the ALP profile on the pion potential becomes negligibly small, and a pure cosine shape is recovered. Correspondingly the pion field $\pi_b(z)$ can only interpolate between $\pi_b(-\infty)=0$ and $\pi_b(+\infty) = 2 \pi - \mathcal{O}(f_a/f_a^\prime)$. This means that the $\pi_b(z)$ profile approaches the sine--Gordon kink with vanishing reflection off the ALP--$\pi_0$ wall, consistently with the ALP decoupling from QCD.}. 

\begin{figure}
    \centering
    \includegraphics[width=0.48\textwidth]{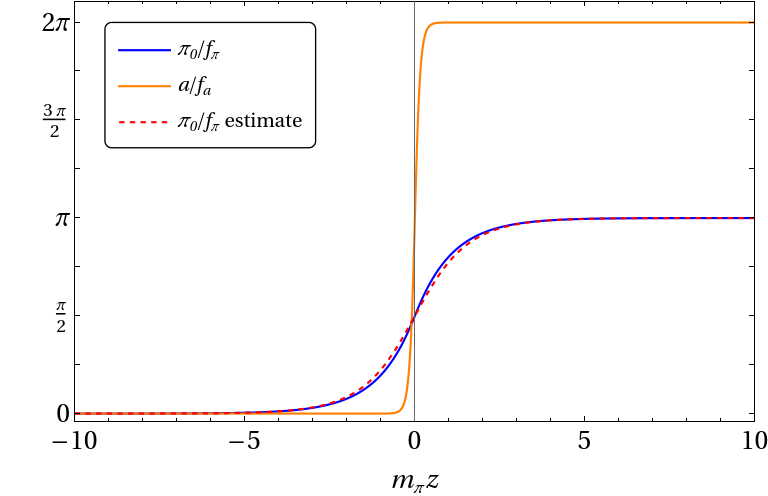}
    \includegraphics[width=0.48\textwidth]{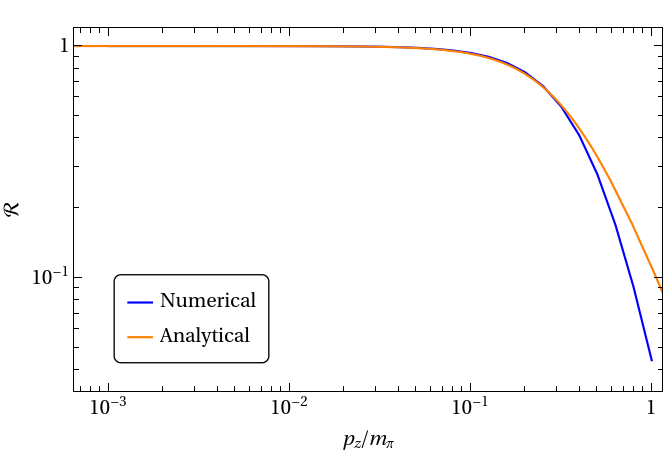}
    \caption{\textbf{Left:} Results for the pion profile $\pi_b(z)$ (blue) in the presence of a realistic ALP background (orange) with $m_\pi/m_a = 0.1$ obtained numerically with a relaxation algorithm. The dashed red line shows the qualitative behavior discussed in the text. The pion profile varies on a scale $\sim m_\pi^{-1}$ much larger than the ALP background $\sim m_a^{-1}$. \textbf{Right} Reflection coefficient evaluated numerically (blue) and the approximation \eqref{eq:Rapp} with $c\simeq 8$ (orange).}
    \label{fig:relax}
\end{figure}

The shape of $\pi_b$ can be obtained by solving numerically the equation of motion. To this end we employ a relaxation algorithm to the extreme approximation of the ALP as a step function as in \eqref{eq:pibminus} and \eqref{eq:pibplus}, as well as to the actual ALP profile with $m_\pi/m_a = 0.1$. We find that as long as the ALP mass is hierarchically larger than $m_{\pi}$, the solution for $\pi_b$ is practically independent of $m_{a}$. 
Qualitatively, one has 
\be 
\label{eq:apppib}
\pi_b(z)/f_\pi \sim 2 \, {\rm arctan} \left(e^{m_\pi z}
\right), \quad m_\pi \ll m_a,
\ee
even though corrections are clearly visible in Fig.\,\ref{fig:relax} (left panel) as \eqref{eq:apppib} does not in fact solve the equation of motion.
Notice that the resulting ALP--$\pi_0$ domain wall has structure at two different scales, $m_a^{-1}$ and $m_\pi^{-1}$, similarly to the case of $\eta^\prime$--$\pi_0$ domain walls in pure QCD\,\cite{Forbes:2000et}.  

Given our background solution, we can study the reflection probability for small oscillations around it. Writing
$\pi_0 = \pi_b(z) + \delta \pi_0(x)$, and $\pi_0(x) = f(z) e^{i E t - i k_x x - i k_y y}$, we have
\be
\label{eq:eomfluctuations}
f''(z) + \left[ k_z^2 + m_\pi^2 - \frac{\partial^2 V}{\partial \pi_0^2}(a(z),\pi_b(z)) \right] f(z) = 0.
\ee
The reflection coefficient is evaluated by solving \eqref{eq:eomfluctuations} numerically. The result is again independent of $m_a$ as long as $m_a \gg m_\pi$,
and it is well approximated by
\be
\label{eq:Rapp}
\mathcal{R}(p_z) \simeq \left(1 + c \frac{p_z^2}{m_\pi^2}\right)^{-1}, \quad p_z \lesssim m_\pi,
\ee
where $c \simeq 8$ (see right panel of Fig.\,\ref{fig:relax}). For $p_z \gtrsim m_\pi$ we are able to identify an exponential drop as expected when the momentum of the scattering particle is of the same order as the inverse wall width $\sim m_\pi^{-1}$. However, this kinematic region is irrelevant for our analysis, as we apply the pion Lagrangian only at temperatures $T \lesssim 60 \, {\rm MeV}$ where high--momentum excitations are Boltzmann suppressed.

Using our result for the reflection coefficient and \eqref{eq:pressurea}, we can straightforwardly evaluate the pressure from pion reflection in the alignment limit:
\be
\Delta P_\pi \simeq v_w g_\pi \frac{1}{8\pi^2} m_\pi^2 \left( m_\pi T + T^2\right) e^{-m_\pi/T}, \quad T < m_\pi \ll m_a,
\ee
where we have included $g_\pi =3$ expecting a similar contribution from the charged pions.

\section{Implications of friction for PTAs}
\label{sec:results}
In this final section
we summarize our results by indicating in the ALP parameter space where deviations from the scaling regime of the DW network are to be expected at temperatures around annihilation,
possibly affecting the SGWB signal.

Notice that for temperatures $ 60 \,{\rm MeV} < T < 2 \,{\rm GeV}$ the pressure from the hadronic sector is not calculable within our simple approach. However, to extract information on this intermediate temperature range, we can look at the pressure at higher temperature (from gluons) and at lower temperature (from pions).
Notice that for $T < 60 \,{\rm MeV}$ we consider the pressure from pions that would act on the domain walls as if they were to be still around (see the previous section for the details of this argument). In practice, the NANOGrav data suggests annihilation temperatures $T_\ast > 60 \,{\rm MeV}$ for our model with the QCD bias, so that the would--be pion pressure is only useful for this extrapolation. For instance, if both the gluon and would--be pion pressure were to dominate in their temperature range of validity, we would conclude that annihilation in the intermediate temperature range is very likely to occur during friction domination.
Otherwise, if friction dominates in only one of the two calculable regimes, a more detailed analysis around the QCD crossover is needed.

\begin{figure}
    \centering
    \includegraphics[width=0.48\textwidth]{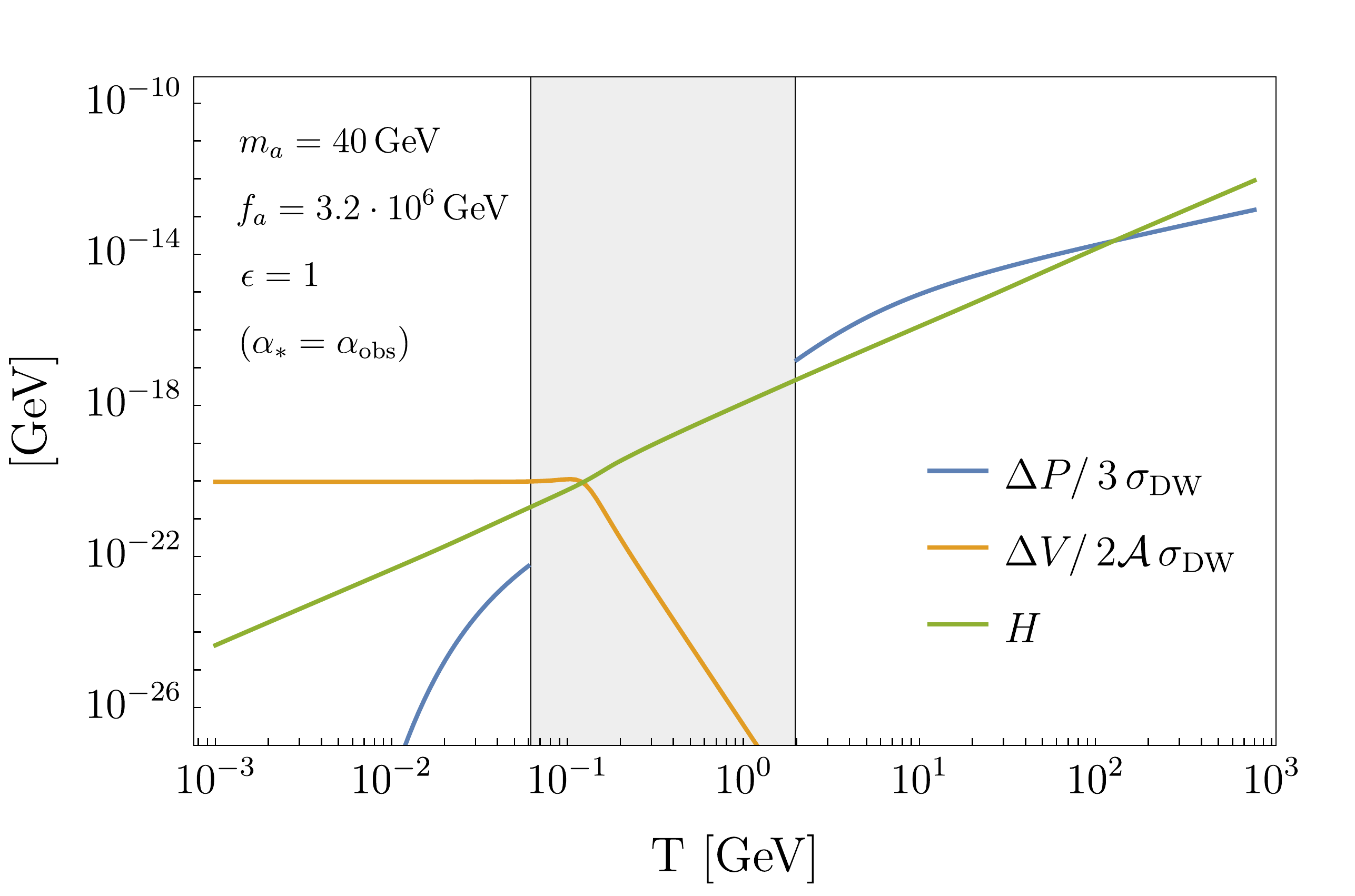}
\includegraphics[width=0.48\textwidth]{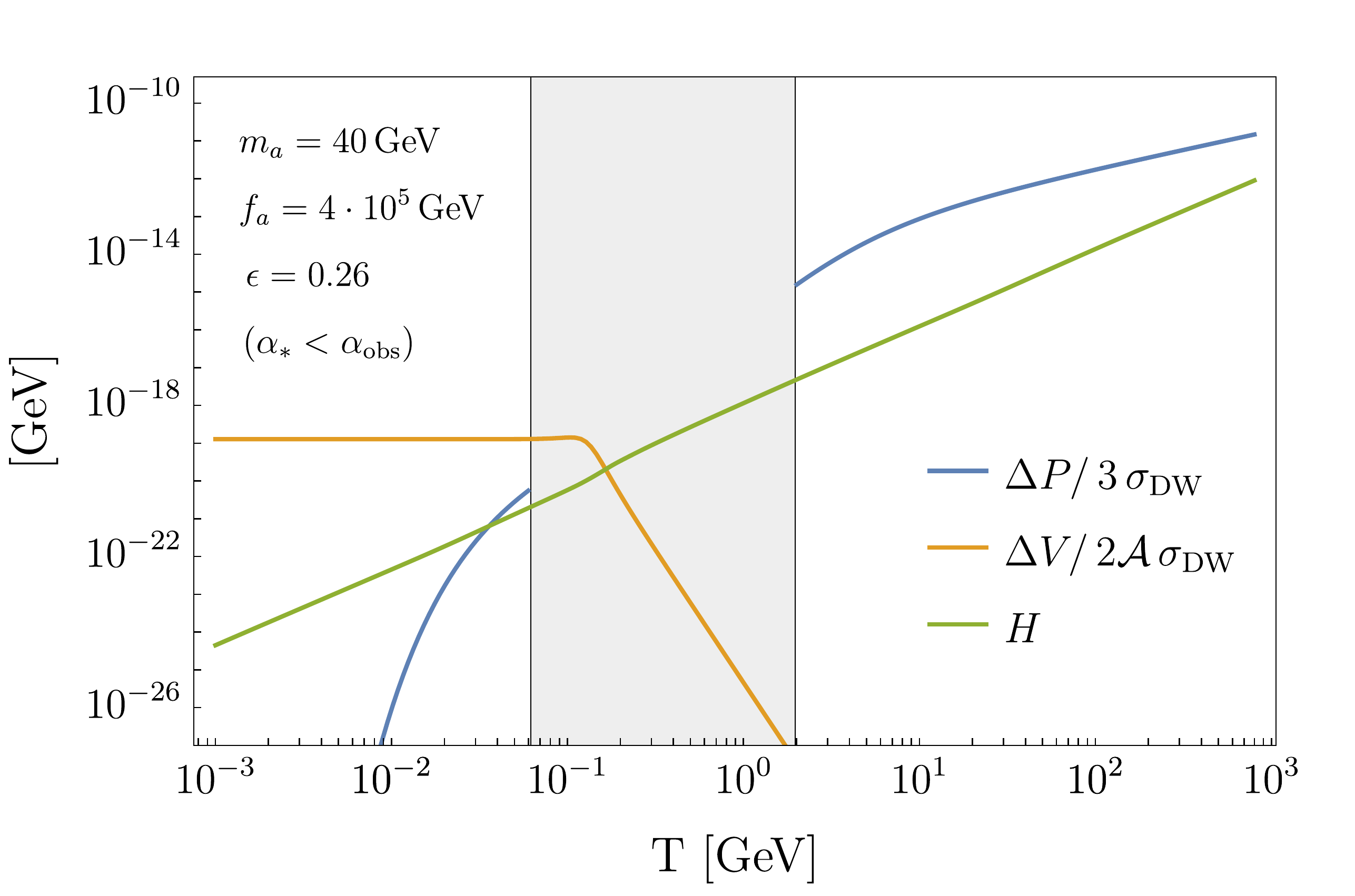}
    \caption{Benchmark points illustrating our friction--domination analysis. \textbf{Left:} Friction domination occurs for temperatures $100 \, {\rm GeV} > T > 2\,\rm GeV$ driven by gluon pressure, as the properly normalized pressure (blue line) overcomes Hubble (green line).
    At lower temperature, the would--be pion pressure is unable to drive friction domination. Whether the network will have enough time to go back to scaling at $T_{\rm ann}$ remains uncertain. If this is the case, this benchmark point is able to explain the PTA signal $(\alpha_\ast = \alpha_{\rm obs})$.
    \textbf{Right:} Gluon and would--be pion pressure are both capable of inducing friction domination, and thus it is very likely that the network never goes back to scaling above the annihilation temperature, identified as the crossing between the properly normalized bias (orange line) and Hubble. Points of this kind, however, require a relatively small domain wall tension and would not be able to explain the GWs observation even if they were to annihilate in the scaling regime ($\alpha_\ast < \alpha_{\rm obs}$). The anomaly coefficients have been chosen as $N_c/N_d= 1.5$. }
    \label{fig:bench}
\end{figure}

\begin{figure}
    \centering
    \includegraphics[width=0.75\textwidth]{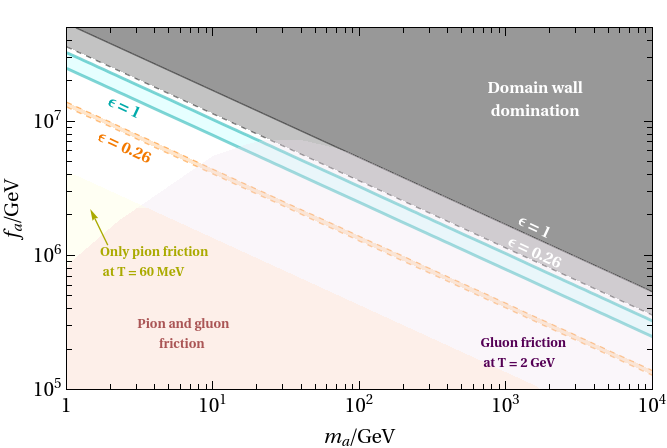}
    \caption{Scan over the $(m_a, f_a)$ parameter space summarizing the results of our analysis. The light blue (orange) band indicates the parameter space that is compatible with the NANOGrav data in the ALP model with a QCD bias considered here with $\epsilon=1$ ($\epsilon=0.26$). As the observed GW background is rather large, both our signal bands are not too far from domain wall domination, shown in the upper right corner by the dark gray (gray) region for $\epsilon=1$ ($\epsilon = 0.26$). 
    The other colored regions highlight the relevance of friction. The purple region corresponds to the parameter space where gluon friction dominates over Hubble at $T=2 \,{\rm GeV}$, where we take $\alpha_s = 0.2$ and $N_c/N_d= 1.5$. This is the lowest temperature where the gluon computation can be trusted, see also Fig.\,\ref{fig:bench}. On the other hand, the would--be pion pressure is evaluated at $T= 60\,{\rm MeV}$ and provides information about the friction in the confined phase, see text for details. The region where the would--be pion pressure can induce friction domination is shown in yellow, and its intersection with the gluon friction region is shown by the pink color.
 The implication for the ALP domain wall interpretation of the PTA data is as follows: for relatively light ALPs with $m_a < 10\,{\rm GeV}$ it is fair to assume that the network annihilates in the scaling regime, so that the signal bands shown here can indeed explain the NANOGrav data. On the other hand, for $m_a > 10\,\rm GeV$ friction is shown to be important at least to the right of the QCD crossover, and a more detailed analysis is required to assess the viability of this interpretation.}
    \label{fig:scan}
\end{figure}

We now illustrate this strategy by 
presenting in Fig.\,\ref{fig:bench} two benchmark points characterized by representative choices of the model parameters. 
The right panel shows a benchmark for which the signal from scaling domain walls and QCD--induced bias can explain the GW signal at PTAs ($\alpha_\ast = \alpha_{\rm obs}$). As we can see, the network does actually enter friction domination around $T \sim 100 \,\rm GeV$ driven by the gluon scattering. Friction remains dominant also at $T \sim 2 \, \rm GeV$, which we take as the edge for the validity of the gluon calculation. 
However, at temperatures $T\sim 60 \, \rm{MeV}$ the would--be pressure from the pions is insufficient to drive the DW network away from scaling. Therefore, it is possible that the period of friction domination ends in the gray region where neither of our calculation is applicable and the DW network goes back to scaling just before annihilation, which in this benchmark is predicted around $T_{\rm ann} \sim 124 \, \rm MeV$\,\footnote{Of course, even if plasma effects become unimportant for $T > T_{\rm ann}$, the network will take a finite time to go back to the scaling regime.}. Points of this type are shown in Fig.\,\ref{fig:scan} in the purple region (labelled by gluon friction), where we suggest that a more refined analysis is needed in order to establish the viability to explain the PTA signal. 

In the right panel of Fig.\,\ref{fig:bench} we show instead a benchmark point for which the gluon and would--be pion are both satisfying the friction domination condition. In this case, it is very likely that the domain walls collapse without ever going back to the scaling regime, with strong implications for the GW signal. However, points of this kind where friction dominates in both our calculable regions require a relatively small tension, and therefore cannot explain the GW signal observed at PTAs even if the network were to annihilate in the scaling regime (emphasized in the right panel as $\alpha_\ast < \alpha_{\rm obs}$). Points of this kind are found in the pion and gluon friction region in Fig.\,\ref{fig:scan}.

Let us now comment on the overall results shown in Fig.\,\ref{fig:scan} as a scan over the $(m_a, f_a)$ parameter space. Additionally to the regions mentioned above, we see that there exists some parameter space for $m_a < 3 \,{\rm GeV}$ where only the would--be pion friction is able to induce fricition domination, while gluons do not. This is understood by noticing that the gluon reflection becomes more and more suppressed as $m_a$ is lowered, see e.g. Eq.\,\eqref{eq:gluonlargema}. On the other hand, as long as $m_\pi \gg m_a$ the would--be pion pressure is independent of $m_a$. This, combined with the fact that gluons need to face a faster Hubble expansion at higher temperatures leads to the only--pion region in Fig.\,\ref{fig:scan}. Notice also that our scan does not extend to points with $m_a < 1 \,{\rm GeV}$ as the approximation $m_a \gg m_\pi$ used in Sec.\,\ref{sec:frictionfrompions} would break down.

Together with the colored regions indicating the impact of friction, we also show in (dark) gray the parameter space where domain walls come to dominate the energy energy of the Universe before annihilation for the choice $\epsilon = 0.26$ ($\epsilon=1$) for the bias in Eq.\,\eqref{eq:bias}. These points are excluded from our analysis. 

The parameter space that can fit the NANOGrav data if the network annihilates in the scaling regime is shown by the light blue band for $\epsilon=1$ and by the narrower orange band for $\epsilon = 0.26$. These signal bands follow straightforwardly from the results shown in Fig.\,\ref{fig:biasvsfit}. As we can see, both these regions are not too far from domain wall domination, as expected given that the preferred values for the network energy density at annihilation are rather large, $\alpha_\star \sim 0.1$. 

The intersection of these signal bands and our friction regions provides the main result of our analysis, which we now summarize. Most of the parameter space compatible with the NANOGrav data implies friction domination from gluons at temperatures $T > 2 \,{\rm GeV}$. However, the would--be pion pressure at low temperatures is not big enough to conclude that the network will be friction dominated at annihilation as well. We nevertheless suggest that a more detailed analysis is needed to ensure viability of these points. On the other hand, for a relatively light ALP with $m_a < 10 \,{\rm GeV}$ we find no evidence for friction domination around the QCD crossover, and thus these points can be viable candidates to explain the PTA data. Even though our analysis cannot extend for $m_a < 1\,{\rm GeV}$, we expect this conclusion to apply also for lighter ALPs.

Before concluding this section, let us mention again that our results only take into account the inevitable friction on the DW network in scenarios with a QCD bias, and that additional, although model--dependent, interactions with the other SM particles can provide important sources of friction as well (see\,\cite{Blasi:2022ayo}).

\section{Conclusions}

The results from PTAs have opened a new era of exploration of the Universe by providing the first evidence of a stochastic background of gravitational waves. 
One possible cosmological explanation for this signal is a network of DWs annihilating around the temperature of the QCD crossover.
An important class of DWs predicted in BSM theories is the one arising in ALP models.
These axionic DWs can accommodate the PTA signal 
if their annihilation is determined by the potential for the ALP that is dynamically induced by QCD as a consequence of the ALP--gluon coupling. However it is important to remark that the prediction of SGWB generated by the DWs is based on numerical simulations that neglect the interaction of the DW with the cosmic plasma, so that the network reaches the scaling regime.

Our main observation is that
in scenarios where the DW annihilation is induced by QCD effects, there is an unavoidable source of friction exerted by QCD states scattering off the DWs.
Our results are summarized in Fig \ref{fig:scan}.
We have identified the portion of the ALP parameter space where friction can be important, even though for domain wall tensions capable of explaining the NANOGrav data we cannot firmly conclude whether friction will be dominant at the annihilation temperature. 
This is because of lack of calculability around the QCD crossover in our simplified approach, and a more refined analysis would be then required.
On the other hand, we were able to identify the region of
ALP parameter space, namely $m_a \lesssim 10 \,\rm{GeV}$,
where friction is negligible and the ALP DW interpretation of the the NANOGrav signal is unaffected.

\section*{Acknowledgments}
All authors are supported in part by the Strategic Research
Program High-Energy Physics of the Research Council
of the Vrije Universiteit Brussel and by the iBOF ``Unlocking the Dark Universe with Gravitational Wave Observations: from Quantum Optics to Quantum Gravity'' of the Vlaamse Interuniversitaire Raad. SB and AM are supported in part by the ``Excellence
of Science - EOS'' - be.h project n.30820817. SB and AR are supported by FWO-Vlaanderen through grant numbers 12B2323N and 1152923N respectively.
AS is supported in part by the FWO-Vlaanderen through the project G006119N. 


 \bibliographystyle{JHEP}
{\footnotesize
\bibliography{biblio}}

\end{document}